\documentstyle[aps,pra,preprint,floats,epsfig]{revtex}
\tightenlines

\def\Red  {}
\def\Black{}
\def\Blue {}
\begin{document}
\draft
\def\fd#1#2{\frac{\displaystyle #1}{\displaystyle #2}}
\def\ket#1{| #1 \rangle}
\def\bra#1{\langle #1 |}
\def\mean#1{\langle #1 \rangle}
\def\Journal#1#2#3#4{{\em #1}, {\bf #2}, #3 (#4)}
\title{\LARGE\bf \Red
Light Pulse Squeezed State Formation In Medium With
The Relaxation Kerr Nonlinearity \Black
\footnote{{\em To appear in}: Proceedings of the 6th
International Conference on Squeezed States and Uncertainty Relations, (ICSSUR'99),
24-29 May 1999, Napoli, Italy, (NASA Conference Publication Series).}}
\author{A.S. Chirkin\footnote{chirkin@foton.ilc.msu.su},
        F. Popescu\footnote{florentin\_p@hotmail.com}}
\address{Faculty of Physics, Moscow State University, Moscow 119899, Russia}
\maketitle
\begin{abstract}\Blue
{\small The consistent theory of formation of pulsed squeezed states as a result of
self-action of ultrashort light pulse in the medium with relaxation Kerr nonlinearity
has been developed. A simple method to form the ultrashort light pulse with
sub-Poissonian photon statistics is analyzed too.}\Black
\end{abstract}
\vspace{7mm}
\section{Introduction}
There are two main approaches in the quantum theory of self-action (or self-phase
modulation) of ultrashort light pulses (USPs). In the first approach (see
\cite{nhh}-\cite{avc}) the calculations of the nonclassical light formation
of the self-action of pulses assume that the nonlinear response of the medium is
instantaneous and that the relative fluctuations are small. The latter assumption is
valid for the intense USPs ordinarily used in experiments. However a finite
relaxation time of the nonlinearity is of principle importance. The relaxation time
of the nonlinearity determines a region of the spectrum of the quantum fluctuations
that play a large role in the formation of squeezed light. In the alternative approach
the inertia of the nonlinearity \cite{bkh,pch} is taken into account.
The methods that
have been developed in \cite{bkh} and \cite{pch} differ in the interaction
Hamiltonian. In \cite{bkh} the authors considered the interaction Hamiltonian
presuming one has to introduce thermal fluctuations in order to satisfy the
commutation relations for time-dependent Bose-operators. For the case of the normally
ordered interaction Hamiltonian \cite{pch} it is not necessary to take into
consideration thermal fluctuations. The results of the quantum theory of the USPs
self-action in the medium with the relaxation Kerr nonlinearity based on the normally
ordered interaction Hamiltonian are presented below. Variances of the quadrature
components and spectral distribution of the pulsed quadrature-squeezed light are
calculated. Besides, propagation of such a pulse in a dispersive linear medium is
analyzed. It is shown that in this case the pulse with sub-Poissonian photon
statistics can be formed.
\section{Quantum theory of the self-action of the light pulse}
\begin{figure}[t]
\vspace{-0.5cm}
    \begin{center}
        \leavevmode
        \epsfxsize=.9\textwidth
        \epsfysize=.51\textwidth
        \epsffile{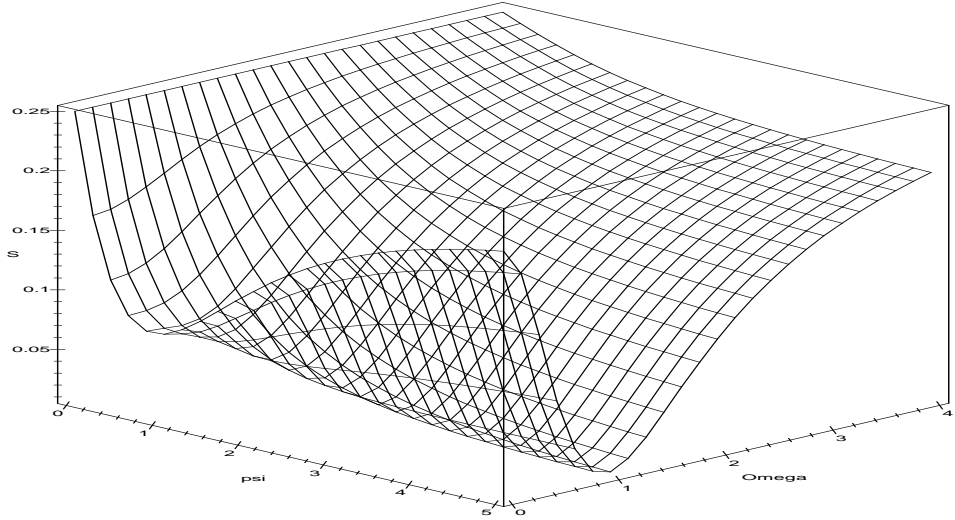}
    \end{center}
    \caption{Spectrum of the fluctuations of the squeezed quadrature (S)
         of a pulse at time t=0 as a function of the maximum nonlinear
         phase $\psi_0$ (psi) and the reduced frequency $\Omega=\omega\tau_r$
         (Omega).\label{fig1}}
\end{figure}

We describe the process under consideration by the following interaction
Hamiltonian
\begin{equation}\label{hamilt}
\hat{H}_{int}(z)=\hbar\beta\int_{-\infty}^{\infty}\,dt\int_{-\infty}^{t}H(t-t_1)
\mbox{\bf\r{N}}[\hat{n}(t,z),\hat{n}(t_1,z)]\,dt_1,
\end{equation}
where the coefficient $\beta$  is determined by the nonlinearity of the medium,
$H(t)$ is the nonlinear response function of the Kerr medium $H(t)\ne 0$ for $t\ge 0$
and $H(t)=0$  for $t<0$; $\mbox{\bf\r{N}}$ is the normal ordering operator,
$\hat{n}(t,z)=\hat{A}^{+}(t,z)\hat{A}(t,z)$ is the photon number density operator,
and $\hat{A}^{+}(t,z)$ and $\hat{A}(t,z)$ are the photon's creation and annihilation
Bose-operators in a given cross section $z$. The operator $\hat{n}(t,z)$ commutes
with the Hamiltonian (\ref{hamilt}) and therefore
$\hat{n}(t,z)=\hat{n}(t,z=0)=\hat{n}_{0}(t)$, where $z=0$ corresponds to the input of
the nonlinear medium. According to Eq.(\ref{hamilt}) the spatial evolution of the
operator $\hat{A}(t,z)$ is given by the equation
\begin{equation}\label{oper}
\frac{\partial\hat{A}(t,z)}{\partial z}-i\beta q[\hat{n}_{0}(t)]\hat{A}(t,z)=0,
\end{equation}
in the moving coordinate system, $z\smash{=}z^{'}$ and
$t\smash{=}t^{'}\smash{-}z^{'}/u$ ($u$ is the velocity of the pulse), and
\begin{equation}\label{qu}
q[\hat{n}_{0}(t)]=\int_{-\infty}^{\infty}h(t_1)\hat{n}_{0}(t-t_1)\,dt_1,\qquad
(h(t)=H(|t|)).
\end{equation}
The solution of Eq.(\ref{oper}) is
\begin{equation}\label{operatorial}
\hat{A}(t,l)=\exp{\{-i\gamma q[\hat{n}_{0}(t)]\}}\cdot\hat{A}_{0}(t).
\end{equation}
Here $\hat{A}_0(t)=\hat{A}(t,0)$, $\gamma=\beta l$, $l$ is the length of the
nonlinear medium. For $h(t)=2\delta(t)$ and $\hat{A}_{0}(t)=\hat{a}_{0}$ expressions
(\ref{qu})-(\ref{operatorial}) have a form corresponding to single-mode radiation. To
verify the commutation relation
$[\hat{A}(t_1,l),\hat{A}^{+}(t_2,l)]=\delta(t_1\smash{-}t_2)$ and to calculate the
quantum characteristics of the pulse it is necessary to develop an algebra of
time-dependent Bose-operators \cite {pch},\cite {blp}. In agreement with
Eq.(\ref{hamilt}) the photon number operator remains unchanged in the nonlinear
medium. This fact has already been used in Eq.(\ref{oper}). Therefore, in the case of
self-action it is of greatest interest to study the fluctuations of the quadrature
components. Here we restrict our consideration by the $X$-quadrature
$\hat{X}(t,z)=[\hat{A}^{+}(t,z)+\hat{A}(t,z)]/2$. The correlation function of the
$X$-quadrature is given by the formula \cite{pch}
\begin{equation}\label{re}
R(t,t+\tau)=\frac{1}{4}\left\{\delta(\tau)-\psi(t)h(\tau)\sin{2\Phi(t)}+\psi^2(t)g(\tau)
\sin^2{\Phi(t)}\right\},
\end{equation}
where $\psi(t)=2\gamma|\alpha_0(t)|^2$ is the nonlinear phase addition, $\alpha_0(t)$
is the eigenvalue of the operator $\hat{A}_0(t)$ of the pulse at the initial coherent
state, $\Phi(t)=\psi(t)+\phi(t)$ ($\phi(t)$ is the initial phase of the pulse). If we
consider the nonlinear response as $h(\tau)=\tau^{-1}_r\exp{(-|\tau|/\tau_r)}$ then
$g(\tau)=\tau^{-1}_r(1+|\tau|/\tau_r)\exp{(-|\tau|/\tau_r)}$ ($\tau_r$ is the
nonlinearity relaxation time). We took into consideration that the parameter $\gamma
\ll 1$ and the pulse duration $\tau_{p}\gg\tau_r$. According to Eq.(\ref{re}), the
spectral density of the quadrature fluctuations is
\begin{equation}\label{spectrum}
S(\omega,t)=\int_{-\infty}^{\infty}R(t,t+\tau)e^{i\omega\tau}\,d\tau=\frac{1}{4}
[1-2\psi(t)L(\omega)\sin2\Phi(t)+4\psi^2(t)L^2(\omega)\sin^2\Phi(t)],
\end{equation}
where $L(\omega)=1/[1+(\omega \tau_r)^2]$.  It follows from Eq.(\ref{spectrum}) that
the level of the quadrature fluctuations, depending on the value of the phase
$\Phi(t)$, can be bigger or smaller than the shot-noise corresponding to
$S^{(coh)}(\omega)=1/4$. If the phase of the pulse is chosen optimal for a frequency
$\omega_0$, $\phi_0(t)=0.5\arctan{\{[\psi(t)L(\omega_0)]^{-1}\}}-\psi(t)$, then the
spectral density at this frequency is minimal. The calculated spectra at $t=0$ for
the case of $\omega_0=\tau^{-1}_r$ are presented in Fig.\ref{fig1}. It is obvious
from Fig.\ref{fig1} that the frequency band in which the spectral density of the
quadrature fluctuations is lower than the shot-noise level depends on the nonlinear
phase addition $\psi(0)$.
\section{Squeezed light pulse in dispersive linear medium}
\begin{figure}[t]
\vspace{-0.5cm}
   \begin{center}
       \leavevmode
       \epsfxsize=.9\textwidth
       \epsfysize=.8\textwidth
       \epsffile{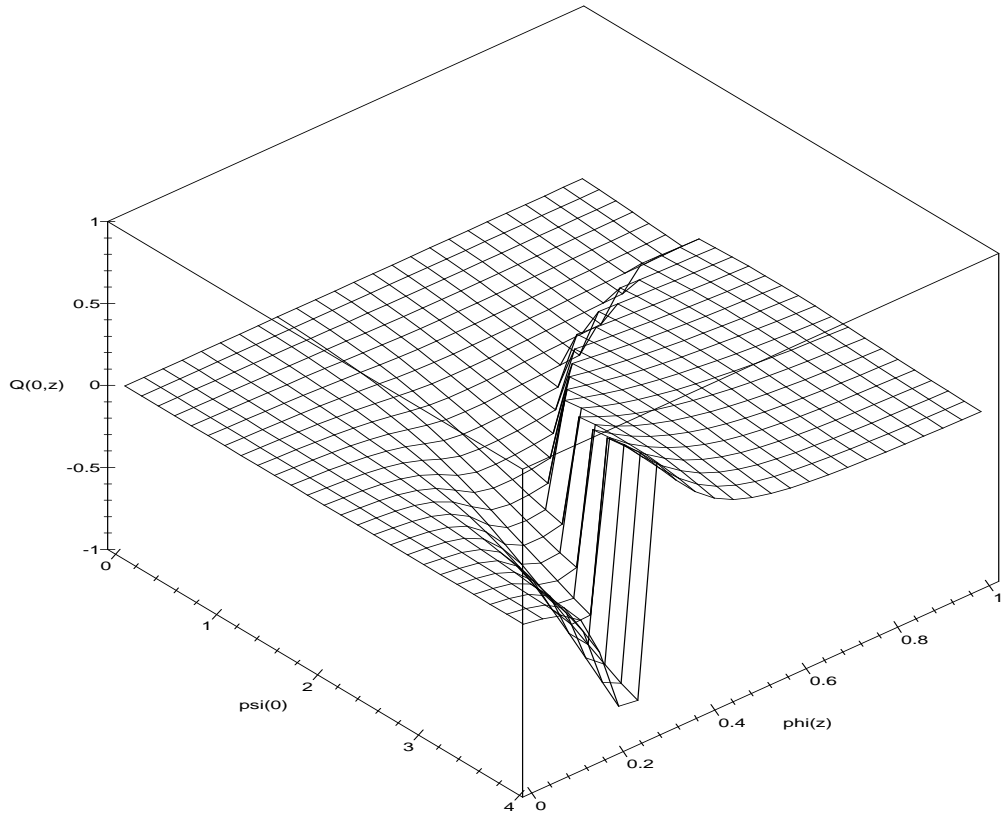}
        \end{center}
        \caption{Parameter $Q(0,z)$ as a function of the maximum nonlinear
        phase $\psi_0$ (psi(0)) and the dispersion phase $\varphi(z)$ (phi(z)).\label{fig2}}
\end{figure}

We analyse now the propagation of the quadrature-squeezed pulse through a dispersive
linear medium in which the following operator transformation take place
\begin{equation}\label{be}
\hat{B}(t,z)=\!\int_{-\infty}^{\infty}\!G(t-t_{1},z)\hat{A}(t_{1},l)dt_{1}.
\end{equation}
Here $G(t,z)$ is the Green function for the medium, $z$ is the distance and
$\hat{A}(t,l)$ is the input value of the operator (at $z=0$) defined by
Eq.(\ref{operatorial}). Let us introduce the photon number operator over the
measurement time $T$ and the Mandel parameter $Q(t,z)$:
\begin{equation}\label{number}
\hat{N}_{T}(t,z)=\!\!\int\limits_{t-T/2}^{t+T/2}\!\!\hat{B}^{+}(t_{1},z)\hat{B}(t_{1},z)dt_{1},
\qquad Q(t,z)=\frac{\varepsilon(t,z)}{\langle\hat{N}_{T}(t,z)\rangle},
\end{equation}
\begin{equation}
\varepsilon(t,z)=\langle\hat{N}^{2}_{T}(t,z)\rangle-
\langle\hat{N}_{T}(t,z)\rangle^{2}-\langle\hat{N}_{T}(t,z)\rangle.
\end{equation}
Let us assume that the initial light pulse has Gaussian form
$\bar{n}_{0}(t)=\bar{n}_0\exp{\{-t^{2}/\tau^{2}_{p}\}}$  and the Green function
equals to $G(t,z)=(-i2\pi k_{2}z)^{-1/2}\cdot\exp{\left\{-it^{2}/2k_{2}z\right\}}$.
The coefficient $k_{2}$ characterizes the dispersion of the group velocity. In the
case of the normal dispersion $k_{2}>0$ and for the anormal dispersion $k_{2}<0$.
When a phase self-modulated pulse passes through the dispersive linear medium the
compression or the decompression of the pulse takes place. This effect can change the
photon statistics of the pulse. In the so-called paraxial approximation we get
\cite{pc}
\begin{equation}
\langle\hat{N}_{T}(t,z)\rangle=\bar{n}_{0}TV^{-1}(z)\exp{[-t^{2}/V^{2}(z)\tau^{2}_{p}]},
\qquad V^{2}(z)=w^{2}(z)+\varphi^{2}(z),
\end{equation}
\begin{equation}\label{quuu}
Q(0,z)=\!-\!\left[\!\frac{T \psi_{0}}{\sqrt{\pi}\tau_{p}}\right]
\!\cdot\!\frac{\sin{[\arctan(\varphi(z)/w(z))+
0.5\arctan{[2\varphi(z)w(z)/(2\varphi(z)\varphi_{d}(z)-w^{2}(z))]]}}}
{[w^{4}(z)-2\varphi^{2}(z)w^{2}(z)+4\varphi^{4}(z)]^{1/4}},
\end{equation}
\begin{equation}
w^{2}(z)=1-s\psi_{0}\varphi(z),\quad\varphi(z)=z/D, \quad \varphi_{d}(z)=z/d, \quad
D=\tau^{2}_{p}/|k_{2}|, \quad d=\tau^{2}_{r}/|k_{2}|.
\end{equation}
$D$ and $d$ are the characteristic dispersion lengths, $s=1$ for $k_{2}<0$, and
$s=-1$ for $k_{2}>0$. It follows from Eq.(\ref{quuu}) that a pulse with
sub-Poissonian photon statistics ($Q(t,z)<0$) can be obtained. Of particular interest
is the compression of the phase self-modulation pulse ($s=1$, $k_{2}<0$). The
dependence of the Mandel parameter in this case is presented in Fig.\ref{fig2}. One
can see that the suppression of quantum fluctuation of the photon number becomes
noticeable for the nonlinear phase $\psi_{0}>1$.
\section{Conclusions}
The main results of the developed systematic theory are as follows. The spectral
region with level of the quadrature fluctuations less than the shot noise depends
on relaxation time of the nonlinearity and the nonlinear phase addition. The choice
of the initial phase of pulse gives us possibility to control the frequency when the
coefficient of squeezing is maximal. The propagation of the quadrature-squeezed light
pulse through a dispersion linear medium (an optical fiber or optical compressor) can
lead to formation of the pulse with sub-Poissonian photon statistics.
\section*{Acknowledgements}
One the authors (A.C.) would like to thank the Organizing Committee of ICSSUR'99  for
financial support for participation in the Conference.

\end{document}